\documentstyle[prd,preprint,tighten,aps,floats,%
amssymb,amsfonts,newlfont,graphicx]{revtex}
\begin{document}
\draft
\preprint{
\begin{tabular}{r}
DFTT 31/00
\\
arXiv:hep-ph/0007155
\end{tabular}
}
\title{A frequentist analysis of solar neutrino data}
\author{M.V. Garzelli and C. Giunti}
\address{INFN, Sez. di Torino, and Dip. di Fisica Teorica,
Univ. di Torino, I--10125 Torino, Italy}
\date{July 13, 2000}
\maketitle
\begin{abstract}
We calculate with Monte Carlo the goodness of fit
and the confidence level of the standard allowed regions
for the neutrino oscillation parameters
obtained from the fit of solar neutrino data.
We show that the values of the
goodness of fit
and of the confidence level of the allowed regions
are significantly smaller than the standard ones.
Using Neyman's method,
we also calculate exact allowed regions
with correct frequentist coverage.
We show that the standard allowed region
around the global minimum of the least-squares function
is a reasonable approximation of the exact one,
whereas the size of the other regions is dramatically
underestimated in the standard method.
\end{abstract}

\section{Introduction}
\label{Introduction}

The standard method to analyze solar neutrino data
in terms of neutrino oscillations
consists in performing a least-squares fit.
However,
for the reasons described in Section~\ref{Standard}
the standard least-squares analysis of solar neutrino data
is approximate from a statistical point of view.

In this paper we present statistical methods
based on Monte Carlo numerical calculations
that allow
to improve the implementation
of the least-squares fit of solar neutrino data.
In Section~\ref{Standard}
we review the standard method
and we discuss why its approximate assumptions could lead
to significant inaccuracy in the results.
In Section~\ref{GOF}
we present a Monte Carlo method
that allows to calculate the goodness of fit
of solar neutrino data.
In Section~\ref{CL}
we present a Monte Carlo method
that allows to calculate the confidence level
of the usual allowed regions in the space of the
neutrino oscillation parameters.
In Section~\ref{Exact}
we present an implementation to solar neutrino analysis
of the classical frequentist Neyman method
that allows to calculate exact confidence regions
with correct coverage.

Since
the purpose of this paper is to illustrate
different methods for the statistical analysis of solar neutrino data,
we consider for simplicity only the data relative to the total rates
measured in the Homestake
\cite{Homestake-98}
and Super-Kamiokande
\cite{SK-sun-lp99}
experiments,
and the weighted average of the total rates measured
in the two Gallium experiments
GALLEX \cite{GALLEX-99}
and SAGE \cite{SAGE-99}.
The values of these rates are given in Table I of Ref.~\cite{Concha-sun-99}.
Updated results of
the Super-Kamiokande experiment
and
first results of the new GNO experiment
have been presented
in the recent Neutrino 2000 conference \cite{nu2000}.
Since the numerical calculations presented here
take a long time
and were started before the
Neutrino 2000 conference,
we do not take into account the new data.
A complete analysis including the new data
and the Super-Kamiokande data relative to the
electron energy spectrum and the zenith-angle distribution
is under way and will be published elsewhere
\cite{Garzelli-Giunti-preparation}.

Neutrino oscillations\footnote{
Here we consider the minimal two-neutrino model,
although more complicated models are possible
(see \cite{Bilenky-Petcov-RMP-87,BGG-review-98}).
}
depend on the mass-squared difference
$\Delta{m}^2 \equiv m_2^2 - m_1^2$
and
on the mixing angle $\vartheta$,
that is restricted in the interval $[0,\pi/2]$.
Traditionally solar neutrino data
have been analyzed in terms of the parameters
$\Delta{m}^2$ and $\sin^22\vartheta$,
that determine the probability of neutrino oscillations in vacuum.
However,
it has recently been shown that
the parameter $\tan^2\vartheta$ is more convenient
for finding the allowed regions in the interval
$\pi/4 \leq \vartheta \leq \pi/2$
when matter effects are important
\cite{Friedland-dark-00,Concha-dark-00}\footnote{
For the same reason
the parameter
$\tan^2 \vartheta$ has been employed
in the framework of three-neutrino mixing
\cite{Fogli-Lisi-Scioscia-95,%
Fogli-Lisi-Montanino-sun-96,%
Fogli-Lisi-Montanino-Scioscia-sacrifice-97}
and the parameter
$\sin^2 \vartheta$ has been employed
in the framework of four-neutrino mixing
\cite{Concha-foursolar-00}.
}.
Moreover,
the parameter
$\tan^2 \vartheta$
allows a better view of the regions at large mixing angles
with respect to the usual parameter $\sin^2 2 \vartheta$.
Hence,
in the following we analyze the solar neutrino data
in terms of the parameters
$\Delta{m}^2$ and $\tan^2 \vartheta$.

Our calculation of the theoretical event rates
follows the standard method described in several papers
for matter-enhanced MSW
\cite{MSW}
transitions
\cite{Kuo-Pantaleone-RMP-89,%
Krastev-Petcov-ANALYTIC-88,%
Krastev-Petcov-unconventional-93}
and
vacuum oscillations
\cite{Krastev-petcov-vo-92,%
Krastev-Petcov-unconventional-93}.
We calculate the MSW survival probability of $\nu_e$'s in the Sun
using the standard analytic prescription
\cite{Parke-86,%
Krastev-Petcov-ANALYTIC-88,%
Kuo-Pantaleone-RMP-89,%
BGG-review-98}
and the level-crossing probability
appropriate for an exponential density profile
\cite{Petcov-analytic-87,Kuo-Pantaleone-RMP-89}.
We calculate the regeneration in the Earth
using a two-step model of the Earth density profile
\cite{Liu-Maris-Petcov-earth1-97,Petcov-diffractive-98,%
Akhmedov-parametric-99,%
Chizhov-Petcov-earth-1-99,Chizhov-Petcov-earth-2-99},
that is known to produce results that do not differ appreciably
from those obtained with the correct density profile.
We have used the tables of neutrino fluxes,
solar density and radiochemical detector cross sections
available in Bahcall's web page \cite{Bahcall-WWW}.
For simplicity
we have neglected the matter effects that slightly affect the
vacuum oscillation solutions of the solar neutrino problem,
as discussed in
\cite{Friedland-vo-00,Fogli-Lisi-Montanino-Palazzo-Quasi-vacuum-00}.

\section{Standard statistical analysis}
\label{Standard}

The traditional way
to find the values of the neutrino oscillation parameters
$\Delta{m}^2$, $\tan^2\vartheta$
allowed by solar neutrino data
is to perform a least-squares fit,
often called ``$\chi^2$ fit''.
In this method the estimates
$\widehat{\Delta{m}^2}$, $\widehat{\tan^2\vartheta}$
of the parameters $\Delta{m}^2$, $\tan^2\vartheta$
are obtained by minimizing the least-squares function
\begin{equation}
X^2
=
\sum_{j_1,j_2}
\left( R^{\mathrm{(thr)}}_{j_1} - R^{\mathrm{(exp)}}_{j_1} \right)
(V^{-1})_{j_1j_2}
\left( R^{\mathrm{(thr)}}_{j_2} - R^{\mathrm{(exp)}}_{j_2} \right)
\,,
\label{X2}
\end{equation}
where $V$ is the covariance matrix of
experimental and theoretical uncertainties,
$R^{\mathrm{(exp)}}_{j}$
is the event rate measured in the $j^{\mathrm{th}}$ experiment
and
$R^{\mathrm{(thr)}}_{j}$
is the corresponding theoretical event rate,
that depends on
$\Delta{m}^2$ and $\tan^2\vartheta$.

The standard method for
the calculation of the covariance matrix $V$
is the one presented in
Refs.~\cite{Fogli-Lisi-correlations-95,Fogli-Lisi-Montanino-Palazzo-3nu-00},
in which the independent uncertainties
$\sigma^2_{j}$
of the experimental rates $R^{\mathrm{(exp)}}_{j}$,
and
the uncertainties of the theoretical rates $R^{\mathrm{(thr)}}_{j}$
are added in quadrature.
Here we use this method,
with the only difference
that we assume a complete correlation of the errors of the
averaged cross sections for the fluxes in each experiment
\cite{Garzelli-Giunti-cs-00}.
Since these correlations are not known,
the choice of complete correlations
is the safest approach.
Hence,
using the notation of
Refs.~\cite{Fogli-Lisi-correlations-95,%
Fogli-Lisi-Montanino-Palazzo-3nu-00}\footnote{
The indices $j,j_1,j_2=1,2,3$ indicate the three solar neutrino experiments
GALLEX+SAGE \cite{GALLEX-99,SAGE-99},
Homestake
\cite{Homestake-98}
and Super-Kamiokande
\cite{SK-sun-lp99},
respectively.
The indices $i,i_1,i_2=1,\ldots,8$
denote the solar neutrino fluxes produced in the eight
solar thermonuclear reactions
$pp$,
$pep$,
$\mathrm{He}p$,
$\mathrm{Be}$,
$\mathrm{B}$,
$\mathrm{N}$,
$\mathrm{O}$,
$\mathrm{F}$,
respectively.
The index $k=1,\ldots,11$
indicate the eleven input astrophysical parameters
in the SSM
(see
Refs.~\cite{Fogli-Lisi-correlations-95,Fogli-Lisi-Montanino-Palazzo-3nu-00}).
},
the covariance matrix $V$ is given by
\begin{equation}
V_{j_1,j_2}
=
\delta_{j_1,j_2}
\sigma^2_{j_1}
+
\delta_{j_1,j_2}
\left(
\sum_{i_1} R_{i_1 j_1}^{\mathrm{(thr)}} \Delta \ln C_{i_1 j_1}^{\mathrm{(thr)}}
\right)^2
+
\sum_{i_1,i_2}
R_{i_1 j_1}^{\mathrm{(thr)}} R_{i_2 j_2}^{\mathrm{(thr)}}
\sum_k
\alpha_{i_1 k}
\alpha_{i_2 k}
\left( \Delta \ln X_k \right)^2
\,,
\label{cov-rat}
\end{equation}
where
\begin{equation}
R_{ij}^{\mathrm{(thr)}}
=
\phi_{i}^{\mathrm{SSM}} \ C_{ij}^{\mathrm{(thr)}}
\label{Rij}
\end{equation}
is the event rate in the $j^{\mathrm{th}}$
experiment
due to the neutrino flux
$\phi_{i}^{\mathrm{SSM}}$
produced in the $i^{\mathrm{th}}$
thermonuclear reaction in the sun
according to the SSM
and
$C_{ij}^{\mathrm{(thr)}}$
is the corresponding energy-averaged cross section
that depends on $\Delta{m}^2$ and $\tan^2\vartheta$.
The quantity
$
\Delta\ln C_{ij}^{\mathrm{(thr)}}
=
\Delta C_{ij}^{\mathrm{(thr)}} / C_{ij}^{\mathrm{(thr)}}
$
is the relative uncertainty
of the energy-averaged cross section $C_{ij}^{\mathrm{(thr)}}$,
that is taken to be approximately equal
to the one calculated without neutrino oscillations.

The quantities $X_k$ are the input astrophysical parameters
in the SSM,
whose relative uncertainties
$\Delta \ln X_k$
determine the correlated uncertainties of the neutrino fluxes
$\phi_{i}^{\mathrm{SSM}}$
through the logarithmic derivatives
\begin{equation}
\alpha_{i k}
=
\frac{\partial\ln\phi_{i}^{\mathrm{SSM}}}{\partial\ln X_k}
\,.
\label{alpha}
\end{equation}
The values of
$\Delta\ln C_{ij}^{\mathrm{(thr)}}$,
$\alpha_{i k}$,
$\Delta \ln X_k$
are given in Ref.~\cite{Fogli-Lisi-Montanino-Palazzo-3nu-00}.

Notice that,
since the theoretical rates
$R_{ij}^{\mathrm{(thr)}}$
depend on $\Delta{m}^2$ and $\tan^2\vartheta$,
also the covariance matrix $V$ depends on
$\Delta{m}^2$ and $\tan^2\vartheta$.

In the traditional method
the minimum $X^2_{\mathrm{min}}$ of (\ref{X2})
provides the estimate of the neutrino oscillation parameters,
usually called ``best-fit values'',
$\widehat{\Delta{m}^2}$ and $\widehat{\tan^2\vartheta}$.
The goodness of the fit is estimated by calculating
the probability to observe a minimum
of $X^2$
larger than the one actually observed
assuming for $X^2_{\mathrm{min}}$
a $\chi^2$ distribution with $N_{\mathrm{exp}}-N_{\mathrm{par}}=1$
degrees of freedom,
where $N_{\mathrm{exp}}=3$ is the number of experimental data points
(the sums over $j_1$ and $j_2$ in Eq.~(\ref{X2})
are from 1 to $N_{\mathrm{exp}}$)
and $N_{\mathrm{par}}=2$ is the number of fitted parameters.
Calling $\alpha$ this probability,
one says that the fit is acceptable at $100\alpha\%$ CL.
If $\alpha$ is larger than a minimum acceptable value,
usually $\sim 10^{-2}$,
the fit is considered to be acceptable
and one can proceed further to determine
the uncertainties in the determination of the parameters
$\Delta{m}^2$ and $\tan^2\vartheta$
(the allowed regions in parameter space).

The standard regions of the parameters allowed at $100\beta\%$ CL
are those that satisfy the condition
\begin{equation}
X^2 = X^2_{\mathrm{min}} + \Delta{X^2}(\beta)
\,,
\label{DX2}
\end{equation}
where
$\Delta{X^2}(\beta)$
is given by the value of $\chi^2$ such that
the cumulative $\chi^2$ distribution for $N_{\mathrm{par}}=2$
degrees of freedom
(the number of parameters)
is equal to $\beta$.
Common values for $\beta$ are
$0.90$ ($ 1.64 \, \sigma$),
$0.95$ ($ 1.96 \, \sigma$),
$0.99$ ($ 2.58 \, \sigma$),
$0.9973$ ($ 3.00 \, \sigma$),
which give
$\Delta{X^2}(0.90) = 4.61$,
$\Delta{X^2}(0.95) = 5.99$,
$\Delta{X^2}(0.99) = 9.21$,
$\Delta{X^2}(0.9973) = 11.83$.

This procedure would be correct if
the theoretical rates
$R^{\mathrm{(thr)}}_{j}$
depended \emph{linearly} on the parameters
$\Delta{m}^2$ and $\tan^2\vartheta$
to be determined in the fit
and the errors
$R^{\mathrm{(thr)}}_{j}-R^{\mathrm{(exp)}}_{j}$
were \emph{multinormally} distributed
with \emph{constant} covariance matrix $V$.
Indeed,
if these requirements were realized
one could prove that
$X^2$
has a $\chi^2$ distribution with $N_{\mathrm{exp}}=3$ degrees of freedom,
$X^2_{\mathrm{min}}$
has a $\chi^2$ distribution with $N_{\mathrm{exp}}-N_{\mathrm{par}}=1$ degrees of freedom,
and
$X^2-X^2_{\mathrm{min}}$
has a $\chi^2$ distribution with $N_{\mathrm{par}}=2$ degrees of freedom
(see \cite{Eadie-71,Frodesen-79,Numerical-Recipes}).
In this case
the $X^2$ function would depend quadratically on the parameters
and there would be only one allowed region
with ellipsoidal form
in the space of the parameters
$\Delta{m}^2$ and $\tan^2\vartheta$.

In the case of solar neutrino data
the gaussian distribution of experimental and theoretical uncertainties
seems to be widely accepted,
although it is not clear if this assumption is appropriate
for the theoretical errors.
On the other hand it is clear that
\begin{enumerate}
\item
The theoretical rates
$R^{\mathrm{(thr)}}_{j}$
do not depend at all linearly on the parameters
$\Delta{m}^2$, $\tan^2\vartheta$.
This is the reason why there are several allowed regions
in the
$\tan^2\vartheta$--$\Delta{m}^2$
plane
(or the more traditional
$\sin^22\vartheta$--$\Delta{m}^2$
plane)
and these regions do not have elliptic form
(see \cite{Bahcall-Krastev-Smirnov-sun-analysis-98,Concha-sun-99}).
\item
The covariance matrix $V$ is not constant,
but depends on $\Delta{m}^2$ and $\tan^2\vartheta$,
as remarked after Eq.~(\ref{alpha}).
\item
The errors
$R^{\mathrm{(thr)}}_{j}-R^{\mathrm{(exp)}}_{j}$
are not multinormally distributed,
because although the fluxes $\phi_i^{\mathrm{SSM}}$
and the cross sections $C_{ij}^{\mathrm{(thr)}}$
are assumed to be multinormally distributed,
their products (\ref{Rij}),
that determine the theoretical rates
through the relations
\begin{equation}
R^{\mathrm{(thr)}}_{j}
=
\sum_i R_{ij}^{\mathrm{(thr)}}
\,,
\label{Rj}
\end{equation}
are not multinormally distributed
(see \cite{Kendall-1}).
\end{enumerate}
Hence, the usual method of calculating
the goodness of fit
and
the allowed regions in the
$\tan^2\vartheta$--$\Delta{m}^2$
plane is not guaranteed to give correct results,
\textit{i.e.}
the goodness of fit could be significantly different
from $100\alpha\%$
and
the confidence level of the regions enclosed by borders with constant
$X^2 = X^2_{\mathrm{min}} + \Delta{X^2}(\beta)$
could be significantly different from $100\beta\%$.

We believe that the largest correction is due to the non-linear
dependence of the theoretical rates
$R^{\mathrm{(thr)}}_{j}$
from the parameters
$\Delta{m}^2$, $\tan^2\vartheta$,
that causes the existence of more than one
local minima of the least-squares function $X^2$.
This implies that there are more possibilities to obtain
good fits of the data and the true goodness of fit
is likely to be smaller than $100\alpha\%$.
Also,
in repeated experiments
the global minimum has significant chances
to occur far from the true (unknown)
value of the parameters
$\Delta{m}^2$, $\tan^2\vartheta$,
with a smaller probability that the allowed regions
cover the true value
with respect to the linear case.
Hence,
we expect that
the true confidence level of a usual
$100\beta\%$ CL allowed region is smaller than $\beta$.

In the following sections
of this paper we perform a least-squares fit of the solar neutrino data
using the $X^2_{\mathrm{min}}$ estimator
for the neutrino oscillation parameters
$\Delta{m}^2$, $\tan^2\vartheta$.
We assume the usual
gaussian distribution
for the experimental and theoretical uncertainties.
In Section~\ref{GOF}
we calculate the goodness of fit
using the Monte Carlo method,
that is applicable
in any case in which the distribution of the uncertainties
is known
(see, for example, Section~15.6 of \cite{Numerical-Recipes}).
In Section~\ref{CL}
we calculate with the Monte Carlo method the confidence level
of the usual allowed regions in the
$\tan^2\vartheta$--$\Delta{m}^2$
parameter space.
In Section~\ref{Exact}
we implement the classical frequentist Neyman method
for finding
exact confidence regions
with correct coverage at a given confidence level.

\section{Goodness of fit}
\label{GOF}

In order to calculate the goodness of fit,
our method proceeds as follows
(see, for example, Section~15.6 of \cite{Numerical-Recipes}).
We estimate the best-fit values of
$\Delta{m}^2$, $\tan^2\vartheta$
through the minimum of $X^2$ in Eq.~(\ref{X2})
and we call these best-fit values
$\widehat{\Delta{m}^2}$, $\widehat{\tan^2\vartheta}$.
Then we assume that
$\widehat{\Delta{m}^2}$, $\widehat{\tan^2\vartheta}$
are reasonable surrogates of the true values
$\Delta{m}^2_{\mathrm{true}}$, $\tan^2\vartheta_{\mathrm{true}}$
and
the probability distribution of the differences
$\widehat{\Delta{m}^2}_{(k)} - \widehat{\Delta{m}^2}$,
$\widehat{\tan^2\vartheta}_{(k)} - \widehat{\tan^2\vartheta}$
is not too different from the true distribution of the differences
$\widehat{\Delta{m}^2}_{(k)} - \Delta{m}^2_{\mathrm{true}}$,
$\widehat{\tan^2\vartheta}_{(k)} - \tan^2\vartheta_{\mathrm{true}}$
in a large set of best-fit parameters
$\widehat{\Delta{m}^2}_{(k)}$, $\widehat{\tan^2\vartheta}_{(k)}$
($k=1,2,\ldots$)
obtained with hypothetical experiments.

Using $\widehat{\Delta{m}^2}$, $\widehat{\tan^2\vartheta}$
as surrogates of the true values,
we generate $N_s$ synthetic random data sets
with the usual gaussian distribution for the
experimental and theoretical uncertainties.
We apply the least-squares method
to each synthetic data set,
leading to an ensemble of simulated best-fit parameters
$\widehat{\Delta{m}^2}_{(s)}$, $\widehat{\tan^2\vartheta}_{(s)}$
with $s=1,\ldots,N_s$,
each one with his associated
$(X^2_{\mathrm{min}})_{s}$.
Then we calculate the goodness of the fit
as the fraction of simulated
$(X^2_{\mathrm{min}})_{s}$
in the ensemble that are larger than the one actually observed,
$X^2_{\mathrm{min}}$.

We calculate the synthetic data sets
generating random neutrino fluxes $\phi_{i}$ with a multinormal distribution
centered on the SSM fluxes $\phi_{i}^{\mathrm{SSM}}$
and having the covariance matrix
\begin{equation}
V^{(\phi)}_{i_1,i_2}
=
\phi_{i_1}^{\mathrm{SSM}}
\phi_{i_2}^{\mathrm{SSM}}
\sum_k
\alpha_{i_1 k}
\alpha_{i_2 k}
\left( \Delta \ln X_k \right)^2
\,.
\label{cov-flux}
\end{equation}
We also generate random energy-averaged cross sections
$C_{ij}$
with a multinormal distribution
centered on the theoretical
energy-averaged cross sections $C_{ij}^{\mathrm{(thr)}}$ corresponding
to
$\widehat{\Delta{m}^2}$, $\widehat{\tan^2\vartheta}$
and having the completely correlated
covariance matrix for each independent experiment $j$
\begin{equation}
V^{(j)}_{i_1,i_2}
=
C_{i_1 j} \Delta \ln C_{i_1 j}
\,
C_{i_2 j} \Delta \ln C_{i_2 j}
\,.
\label{cov-cs}
\end{equation}
Then, we calculate the rates
$
R_j
=
\sum_i \phi_i C_{ij}
$.
Finally,
we generate random synthetic experimental rates
$R_j^{(s)}$
with normal distribution centered on $R_j$
and standard deviation equal to that of the actual experimental data
($\sigma_j$).
The synthetic experimental rates are inserted
in the least-squares function (\ref{X2})
in place of
$R^{\mathrm{(exp)}}_{j}$
in order to find the minimum
$(X^2_{\mathrm{min}})_{s}$
and its associated best-fit parameters
$\widehat{\Delta{m}^2}_{(s)}$, $\widehat{\tan^2\vartheta}_{(s)}$.

The results of our calculations are reported in Table~\ref{gof}.
The global minimum of the least-squares function (\ref{X2}),
$X^2_{\mathrm{min}} = 0.42$,
occurs in the SMA region\footnote{
Here we use the standard terminology for the allowed regions
(see \cite{Bahcall-Krastev-Smirnov-sun-analysis-98,Concha-sun-99}):
SMA for
$\Delta{m}^2 \sim 5 \times 10^{-6} \, \mathrm{eV}^2$,
$\tan^2 \vartheta \sim 10^{-3}$,
LMA for
$\Delta{m}^2 \sim 3 \times 10^{-5} \, \mathrm{eV}^2$,
$\tan^2 \vartheta \sim 0.3$,
LOW for
$\Delta{m}^2 \sim 10^{-7} \, \mathrm{eV}^2$,
$\tan^2 \vartheta \sim 0.5$,
VO for
$\Delta{m}^2 \lesssim 10^{-8} \, \mathrm{eV}^2$.
}
for
$\Delta{m}^2 = 5.1 \times 10^{-6} \, \mathrm{eV}^2$
and
$\tan^2 \vartheta = 1.6 \times 10^{-3}$.
The results reported in the ``SMA'' row of Table~\ref{gof}
have been obtained taking
$\widehat{\Delta{m}^2} = 5.1 \times 10^{-6} \, \mathrm{eV}^2$
and
$\widehat{\tan^2\vartheta} = 1.6 \times 10^{-3}$.
We first restricted the allowed region of the mixing parameters
around the SMA region
($10^{-4} \leq \tan^2\vartheta \leq 3 \times 10^{-2}$
and
$3 \times 10^{-7} \, \mathrm{eV}^2
\leq \Delta{m}^2 \leq
10^{-4} \, \mathrm{eV}^2$)
and obtained the local value of the goodness of fit,
reported in the ``local'' column of Table~\ref{gof}.
This value is almost equal
(even slightly larger)
to the standard one
obtained assuming a $\chi^2$ distribution with one degree of freedom,
reported in the ``standard GOF'' column of Table~\ref{gof}.
Hence, we conclude that locally the usual method to evaluate the
goodness of fit is reliable.

\begin{table}[t!]
\begin{center}
\includegraphics[bb=85 560 460 775]{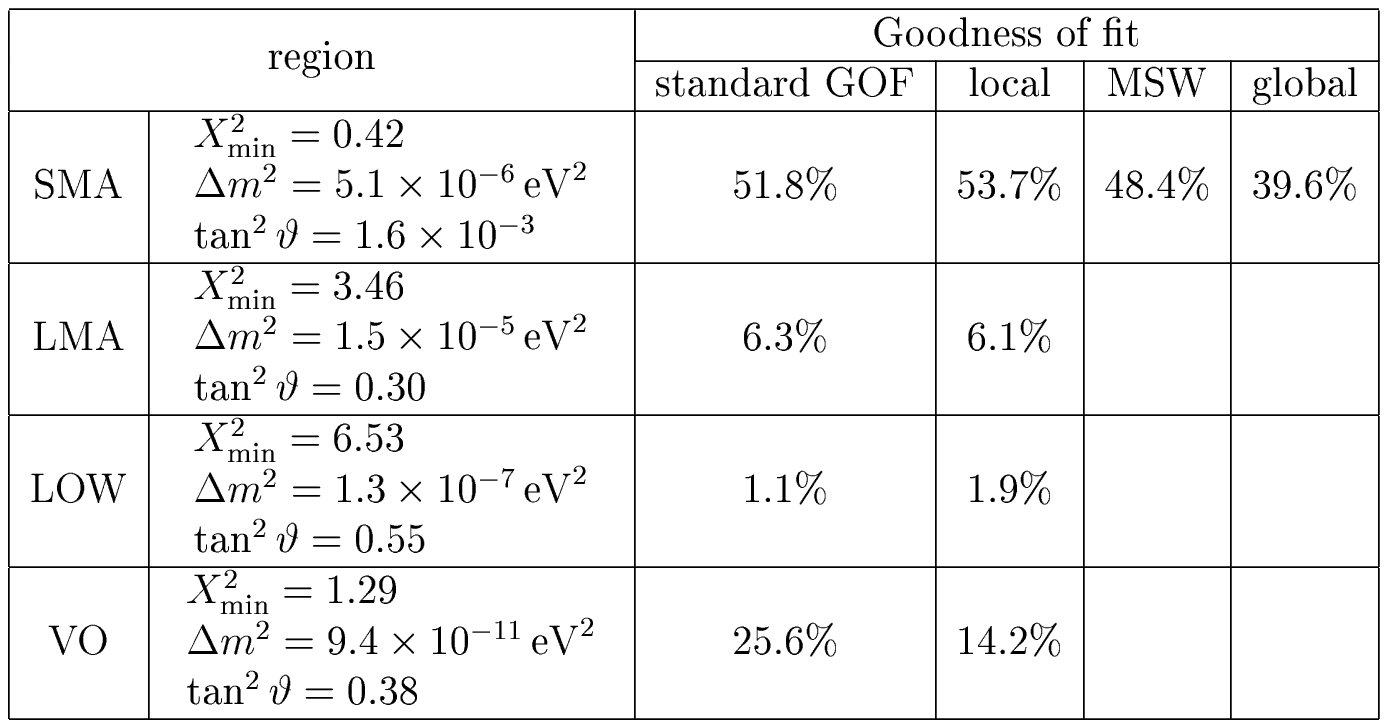}
\end{center}
\caption{ \label{gof}
Goodness of fit of solar neutrino data calculated with
more than one million synthetic data sets.
The first two columns indicate in which region
the surrogate of the true values of the neutrino oscillation parameters
has been assumed to be,
the corresponding values of $X^2_{\mathrm{min}}$
and the values of the surrogates.
The third column indicates the goodness of fit
calculated with the standard method,
\textit{i.e.} assuming a $\chi^2$ distribution
with one degree of freedom.
The fourth column reports the goodness of fit calculated locally,
\textit{i.e.}
restricting the allowed values of the parameters around the
region in which the assumed surrogates of the true values lie.
The fifth column reports the goodness of fit calculated
restricting the allowed values of the parameters to the MSW region
(\protect\ref{MSW}).
The sixth column reports the goodness of fit calculated
without any
restriction on the allowed values of the parameters.
}
\end{table}

However,
when we extend the allowed region of the mixing parameters to all the MSW
region
\begin{equation}
10^{-4} \leq \tan^2\vartheta \leq 2
\,,
\quad
10^{-8} \, \mathrm{eV}^2
\leq \Delta{m}^2 \leq
10^{-3} \, \mathrm{eV}^2
\quad
\mbox{(MSW region)}
\,,
\label{MSW}
\end{equation}
and when we add also the VO region
\begin{equation}
0.1 \leq \tan^2\vartheta \leq 1
\,,
\quad
10^{-11} \, \mathrm{eV}^2
\leq \Delta{m}^2 \leq
10^{-8} \, \mathrm{eV}^2
\quad
\mbox{(VO region)}
\,,
\label{VO}
\end{equation}
we obtain the values reported,
respectively,
in the ``MSW'' and ``global'' columns of Table~\ref{gof},
which are significantly smaller
than the one obtained with the standard method.
As remarked in Section~\ref{Standard},
this is due to the non-linear
dependence of the theoretical rates
from the neutrino oscillation parameters,
that implies that there are more possibilities to obtain
good fits of the data with respect to the linear case.
Therefore,
we conclude that the standard method, although valid locally
(when the allowed region of the parameters is restricted
around the SMA region the linear assumption is approximately correct),
is not valid in general
and should not be trusted if there is more than one allowed region.

In order to check the local validity of the standard method
we have also assumed that
$\widehat{\Delta{m}^2}$
and
$\widehat{\tan^2\vartheta}$
have the values corresponding to the local minima of $X^2$
in the LMA, LOW and VO regions,
restricting the allowed values of the parameters
around the corresponding regions.
The results are reported in the ``LMA'', ``LOW'' and ``VO'' rows
of Table~\ref{gof}.
One can see that the standard method is locally acceptable
for the LMA and LOW solutions,
but it largely overestimates the goodness of fit in
the case of the VO solution.
This is due to the fact that
the theoretical rates are highly non-linear functions
of the neutrino oscillation parameters
in the VO region (\ref{VO}),
resulting in
several disjointed allowed regions.

The ``MSW'' and ``global'' entries in the ``LMA'', ``LOW'' and ``VO''
rows of Table~\ref{gof}
are empty because it is meaningless
to calculate the goodness of fit
allowing values of the parameters
in which the fit is better than the one
in the assumed surrogate of the true values of the parameters.

Summarizing the results of this section,
we have shown that if there were only one allowed region
in the space of the neutrino oscillation parameters,
or if there are valid reasons to restrict the
allowed region of the parameters around
one of the SMA, LMA, LOW solutions,
the standard method to calculate the goodness of fit
is approximately reliable.
On the other hand,
if there are more than one allowed regions,
the standard method to calculate the goodness of fit
is not reliable and the goodness of fit must be calculated numerically,
with Monte Carlo,
as we have done.
This happens if one considers the MSW region (\ref{MSW})
of the neutrino oscillation parameters,
which contains three allowed regions (SMA, LMA and LOW),
or the VO region  (\ref{VO}),
that contains several allowed regions,
or all the parameter space (MSW+VO).

\section{Confidence level of allowed regions}
\label{CL}

In order to calculate the confidence level of the allowed regions
it is necessary first to understand what is its meaning.
The $100\beta\%$ CL allowed regions
are defined by the property that they belong to a
set of allowed regions obtained with hypothetical experiments
and the regions belonging to this set
cover (\textit{i.e.} include) the true value of the parameters
with probability $\beta$.

Given the usual ``$100\beta\%$ CL'' allowed regions
in the space of the neutrino oscillation parameters
we can calculate their confidence level
$\beta_{\mathrm{MC}}$
with a method similar to the one described in the previous section
for the goodness of fit.
We assume that
$\widehat{\Delta{m}^2}$, $\widehat{\tan^2\vartheta}$
are reasonable surrogates of the true values
$\Delta{m}^2_{\mathrm{true}}$, $\tan^2\vartheta_{\mathrm{true}}$
and we
generate a large number of synthetic data sets.
We apply the standard procedure to each synthetic data set
and obtain the corresponding ``$100\beta\%$ CL'' allowed regions
in the space of the neutrino oscillation parameters.
Then we count the number of synthetic
``$100\beta\%$ CL'' allowed regions
that cover the assumed surrogate
$\widehat{\Delta{m}^2}$, $\widehat{\tan^2\vartheta}$
of the true values.
The ratio of this number and the total number of synthetically
generated data set gives the confidence level
$\beta_{\mathrm{MC}}$
of the ``$100\beta\%$ CL'' allowed regions.

The results of our calculations
are reported in Table~\ref{cl}.
As we have done in the previous section for the goodness of fit,
we calculated first the local confidence levels
restricting the allowed values of the parameters around the
region whose local minimum of $X^2$
gives the assumed surrogates of the true values
(``local'' column of Table~\ref{cl}).
Then we calculated the confidence levels
restricting the allowed values of the parameters to the MSW region (\ref{MSW})
assuming the surrogates of the true values in the local minima of $X^2$
of the SMA, LMA and LOW regions
(``MSW'' column of Table~\ref{cl}).
Finally,
we calculated the confidence levels
without any
restriction on the allowed values of the parameters,
assuming the surrogates of the true values in the local minima of $X^2$
of the SMA, LMA, LOW and VO regions
(``global'' column of Table~\ref{cl}).

\begin{table}[t!]
\begin{center}
\includegraphics[bb=85 505 500 775]{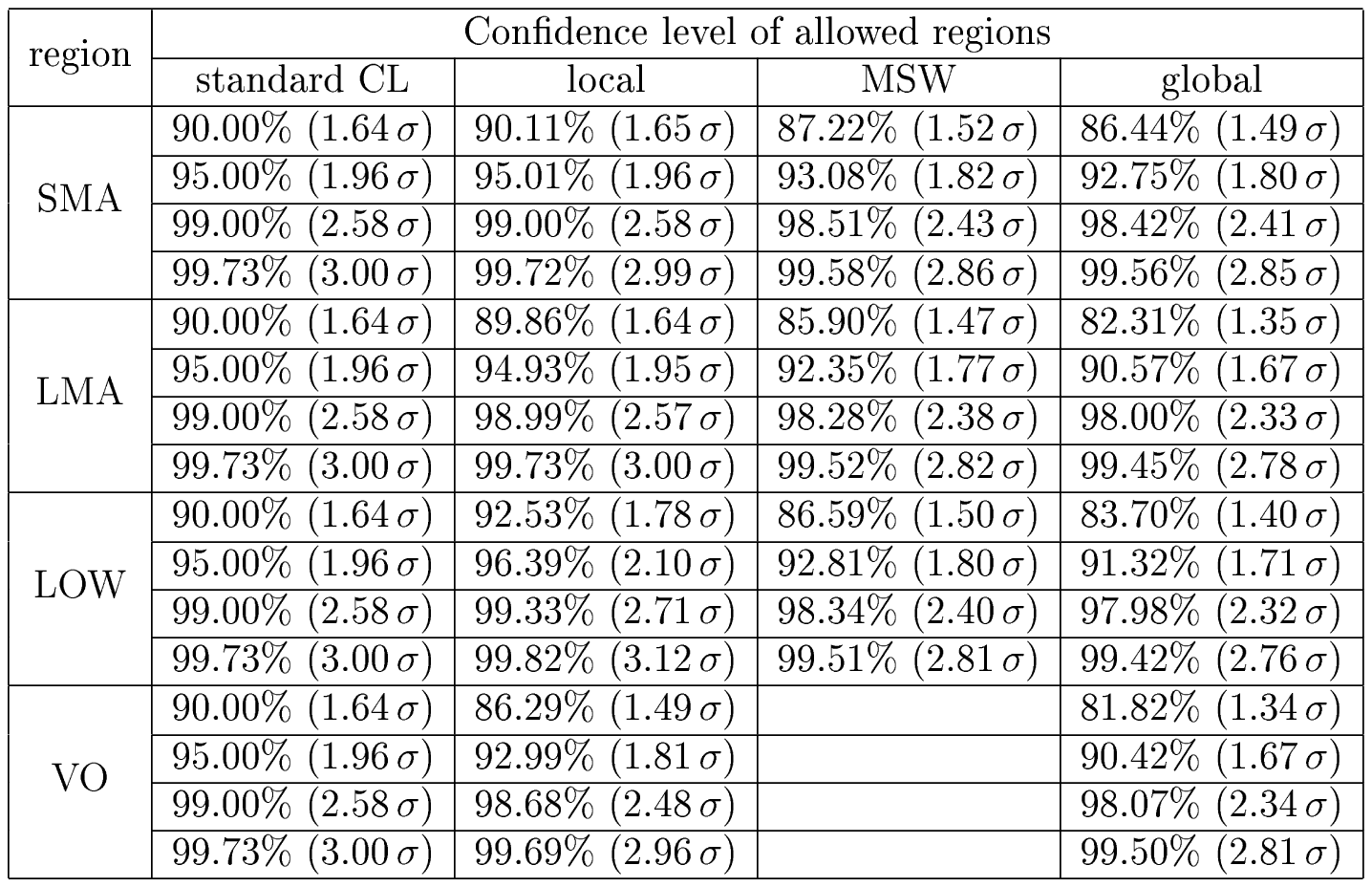}
\end{center}
\caption{ \label{cl}
Confidence level of the usual
90\%, 95\%, 99\% and 99.73\% CL allowed regions.
The confidence levels have been calculated generating
more than one million synthetic data sets.
The first column indicates in which region
the surrogate of the true values of the neutrino oscillation parameters
has been assumed to be.
The second column indicates the usual CL.
The third column reports the confidence levels calculated locally,
\textit{i.e.}
restricting the allowed values of the parameters around the
region in which the assumed surrogates of the true values lie.
The fourth column reports the confidence levels calculated
restricting the allowed values of the parameters to the MSW region
(\protect\ref{MSW}).
The fifth column reports the confidence levels calculated
without any
restriction on the allowed values of the parameters.
}
\end{table}

One can see that the values of the confidence levels
calculated locally
for the SMA region,
where the global minimum of $X^2$ lies,
practically coincide with the standard ones
(``standard CL'' column of Table~\ref{cl}).
However,
when the allowed values of the parameters
are extended to the whole MSW region (\ref{MSW})
or
to the MSW and VO regions (global),
the confidence levels are significantly smaller than
the standard ones.

The same trend, slightly more pronounced, is observed
when the surrogates of the true values of the parameters
are assumed to correspond to the local minima
of $X^2$
in the LMA and LOW region,
with even a small deviation
of the local confidence levels from the standard ones
(with unpredictable sign).

When the surrogates of the true values of the parameters
are assumed to correspond to the local minimum
of $X^2$
in the VO region (\ref{VO}),
the confidence levels are significantly smaller than
the standard ones, even those calculated locally.
This is due to the fact that
the linear approximation used in the
calculation of the standard confidence levels
is badly violated
(there are several disjointed allowed VO regions
with non-elliptical shapes).

Summarizing the results of this section,
we have shown that the standard confidence levels
of the allowed regions in the neutrino oscillation parameter space
are approximately correct
if only one of the SMA, LMA or LOW region is considered to be allowed
a priori.
If the oscillation parameters
are restricted to the MSW region (\ref{MSW}),
the confidence levels are significantly smaller than the standard ones,
with some uncertainty depending on the assumed
surrogates of the true values of the parameters.
If one does not impose any restriction on the values of the parameters,
the confidence levels decrease further.
If only the VO regions are considered to be allowed
even the confidence levels
calculated locally are significantly smaller than the standard ones.

\section{Exact allowed regions}
\label{Exact}

In the previous section we have calculated the confidence level
of the allowed regions in the neutrino oscillation parameter space
obtained with the standard procedure based on Eq.~(\ref{DX2}).
This calculation is approximate,
because it is based on the assumption of a surrogate
for the unknown true values of the neutrino oscillation parameters.
Furthermore,
we have seen that the value of the confidence level
is different if the surrogate
for the unknown true values of the neutrino oscillation parameters
is assumed to be the value of the parameters in the global minimum of $X^2$
or in one of the local minima.

Luckily, there is a well-known procedure for
constructing exact \emph{confidence intervals}
independently of the true values of the parameters.
This procedure has been invented by Neyman in 1937 \cite{Neyman-37}
(see also \cite{Kendall-2A,Eadie-71,PDG-98}).
It guarantees that the resulting confidence intervals have
correct frequentist \emph{coverage}
(see \cite{Cousins-95,Feldman-Cousins-98,%
Giunti-back-99,Giunti-Laveder-00,Giunti-clw-00}),
\textit{i.e.}
they belong to a set of confidence intervals
obtained with different or similar, real or hypothetical experiments
that cover the true values of the parameters
with the desired probability given by the chosen confidence level.
In this section we apply this method in order to find
confidence intervals with proper coverage
for the neutrino oscillation parameters.

Neyman's construction of exact frequentist confidence interval
with $100{\beta}\%$ confidence level
starts with the choice of an appropriate estimator of the parameters
under investigation.
Then,
for any possible value of the parameters
one calculates an \emph{acceptance interval} with probability $\beta$,
\textit{i.e.}
an interval of the estimator
that contains $100{\beta}\%$ of the values of the estimator
obtained in a large series of trials.
Several methods are available for the construction of
the acceptance intervals
(see \cite{Kendall-2A,Eadie-71,PDG-98,Feldman-Cousins-98,%
Giunti-Laveder-00,Giunti-clw-00}
and references therein).
If the probability distribution of the estimator is known,
the acceptance intervals can be calculated analytically;
if not, one can calculate the acceptance intervals
with numerical Monte Carlo methods.
In general the acceptance intervals can be composed by disjoint
sub-intervals.
In the case of $n$ parameters
the acceptance intervals are regions
in the $n$-dimensional parameter space.

Once the $100{\beta}\%$ acceptance interval
for each possible value of the parameters
is calculated,
the $100{\beta}\%$ confidence interval
is simply composed by all the parameter values
whose acceptance interval covers the measured value of the estimator
(\textit{i.e.} the actual estimate of the parameters).
If the acceptance intervals are composed by disjoint
sub-intervals,
also the confidence interval
is composed by disjoint
sub-intervals.
As we will see in the following,
this is what happens in the case of
solar neutrino oscillations.

Our implementation of Neyman's construction goes as follows.
First we choose as estimator of neutrino oscillation parameters
the values of the parameters in the minimum $X^2_{\mathrm{min}}$
of the least-squares function (\ref{X2}).
Since the probability distribution of the chosen estimator is not known,
we calculate it numerically with a Monte Carlo.
We define an appropriate grid in the 2-dimensional space of
the neutrino oscillation parameters
$\tan^2\vartheta$, $\Delta{m}^2$
and for each value of the parameters on the grid
we generate a large number of synthetic data sets.
For each data set we find the value of the parameters
corresponding to the minimum of $X^2$.
This procedure gives the distribution of
$X^2_{\mathrm{min}}$
for each value of the parameters on the grid.
Unfortunately this is a rather lengthy task
that requires several days of computer time
in order to reach an acceptable accuracy,
essentially because of the large number of points
on a reasonably fine grid,
about five thousand in the MSW region (\ref{MSW})
and six thousand in the VO region (\ref{VO}).

We define the
$100{\beta}\%$ acceptance intervals
in the simplest and most natural way\footnote{
There is a subtle problem in choosing the method
that defines the acceptance intervals:
the method must be chosen independently of the data
and the result.
This is what we have done.
Otherwise,
the property of coverage is lost
(see \cite{Cousins-95,PDG-98,Feldman-Cousins-98,%
Giunti-back-99,Giunti-Laveder-00,Giunti-clw-00}),
and one can always choose a method ``ad hoc''
to obtain any desired result.
}:
for each value of the parameters
we choose the shortest possible acceptance interval,
\textit{i.e.}
that containing the values of the parameters on the grid
with highest probability,
whose sum is equal or larger than $\beta$
(in general perfect equality is not reached
because of the discrete nature of the grid).
In the case of a linear least-squares fit
this method gives the allowed regions obtained with
the standard prescription (\ref{DX2}).
Therefore,
our exact allowed regions can be compared directly with the standard ones.

The acceptance intervals are 2-dimensional regions
in the $\tan^2\vartheta$--$\Delta{m}^2$ parameter space.
Because of the non-linearity of the neutrino oscillation
probability as a function of the parameters,
the acceptance intervals are composed by disjoint
sub-intervals.
This generates 2-dimensional confidence intervals
composed by disjoint
sub-intervals,
some of which far from the values
of the parameters corresponding
to the actual $X^2_{\mathrm{min}}$.
The confidence intervals
are composed by the values of the parameters
whose acceptance interval includes
the parameters corresponding to
the actual $X^2_{\mathrm{min}}$
(the measured value of the estimator).

\begin{figure}[t!]
\begin{center}
\includegraphics[height=8cm]{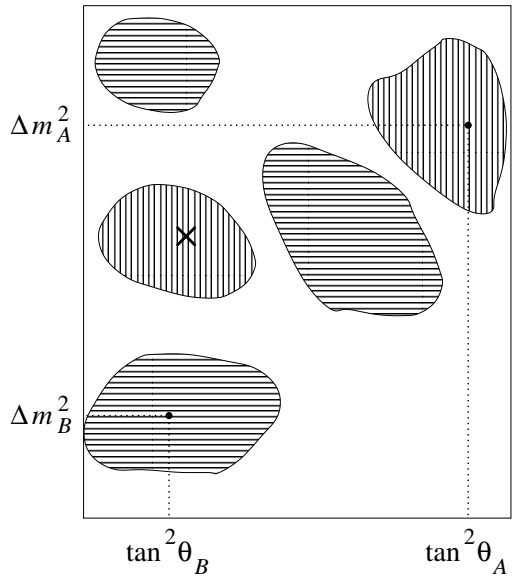}
\end{center}
\caption{ \label{ai}
Illustration of the acceptance intervals.
The cross corresponds to the actual $X^2_{\mathrm{min}}$.
The two vertically hatched regions constitute the acceptance interval
associated with
$\tan^2\vartheta_A$, $\Delta{m}^2_A$.
The three horizontally hatched regions constitute the acceptance interval
associated with
$\tan^2\vartheta_B$, $\Delta{m}^2_B$.
}
\end{figure}

The procedure is illustrated qualitatively in Fig.~\ref{ai},
where
the cross corresponds to the actually measured
$X^2_{\mathrm{min}}$,
the union of the two vertically hatched regions is the acceptance interval
associated with
$\tan^2\vartheta_A$, $\Delta{m}^2_A$,
and
the union of the three horizontally hatched regions is the acceptance interval
associated with
$\tan^2\vartheta_B$, $\Delta{m}^2_B$.
Since the acceptance interval
associated with
$\tan^2\vartheta_A$, $\Delta{m}^2_A$
includes the point corresponding to $X^2_{\mathrm{min}}$,
the point
$\tan^2\vartheta_A$, $\Delta{m}^2_A$
belongs to the confidence interval.
On the other hand,
the acceptance interval
associated with
$\tan^2\vartheta_B$, $\Delta{m}^2_B$
does not include the point corresponding to $X^2_{\mathrm{min}}$
and
the point
$\tan^2\vartheta_A$, $\Delta{m}^2_A$
is out of the confidence interval.

\begin{figure}[p]
\begin{center}
\includegraphics[bb=45 110 540 757,height=0.8\textheight]{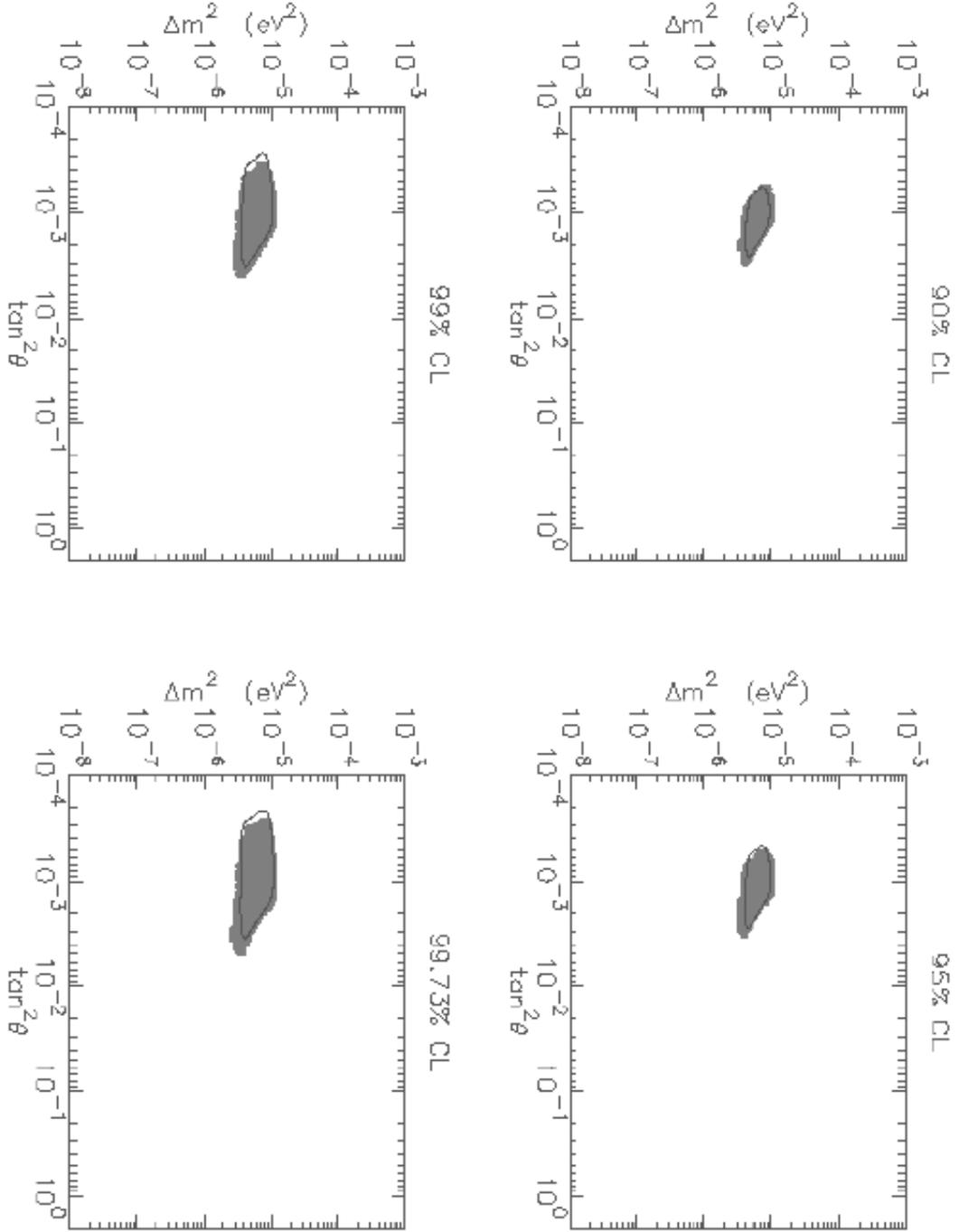}
\end{center}
\caption{ \label{cfi-sma}
Allowed 90\%, 95\%, 99\%, 99.73\% confidence level regions
in the
$\tan^2\vartheta$--$\Delta{m}^2$
plane.
In each plot
the gray area is the allowed region with exact frequentist coverage
obtained restricting the possible
values of $\tan^2\vartheta$ and $\Delta{m}^2$
in a region around the SMA solution,
where $X^2_{\mathrm{min}}$ lies.
The area enclosed by the solid line
is the standard SMA allowed region.
}
\end{figure}

\begin{figure}[p]
\begin{center}
\includegraphics[bb=45 110 540 757,height=0.8\textheight]{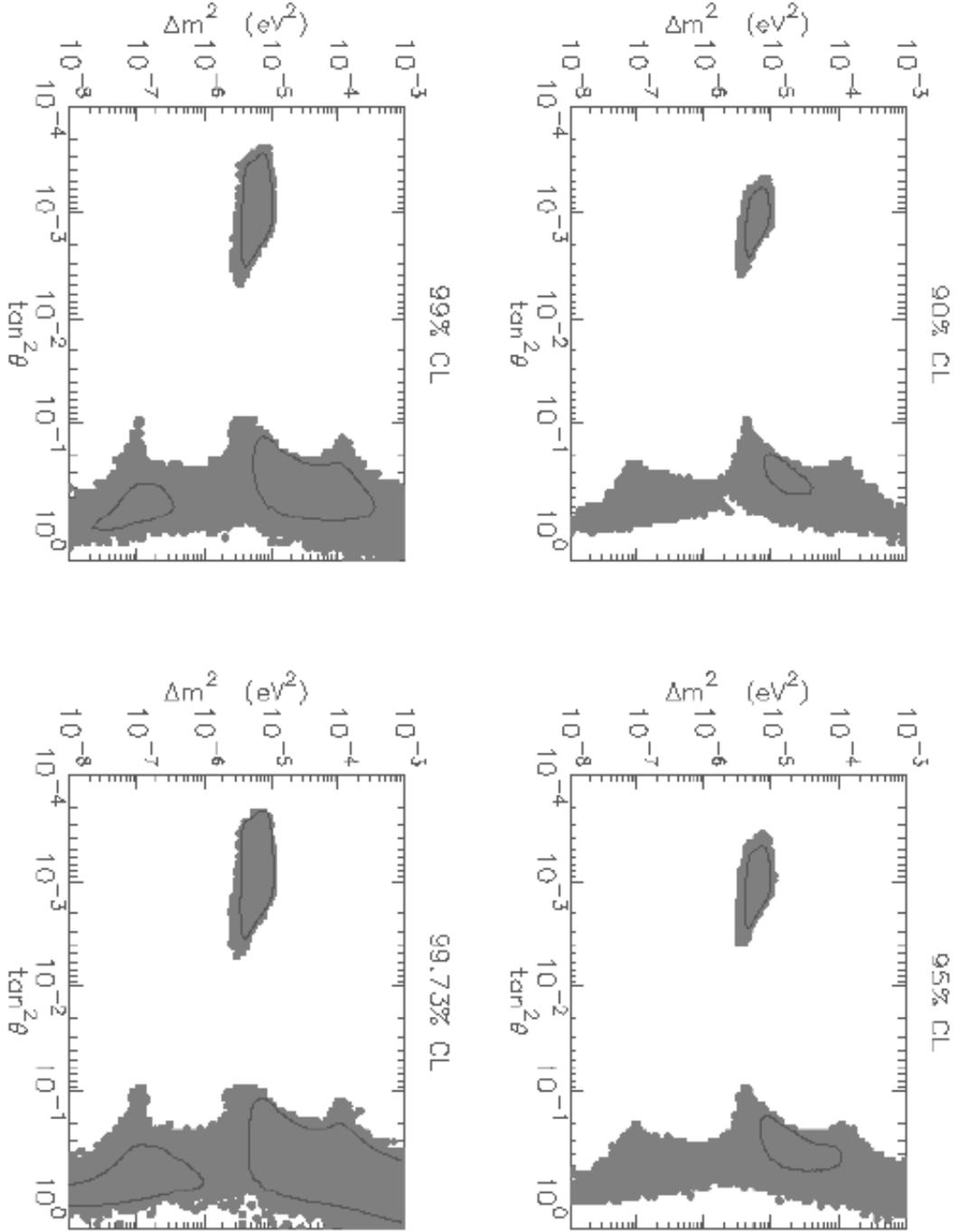}
\end{center}
\caption{ \label{cfi-msw}
Allowed 90\%, 95\%, 99\%, 99.73\% confidence level regions
in the
$\tan^2\vartheta$--$\Delta{m}^2$
plane.
The gray areas are the allowed regions with exact frequentist coverage
obtained considering all possible
values of $\tan^2\vartheta$ and $\Delta{m}^2$
in the MSW region (the whole area of the plots).
The areas enclosed by the solid lines
are the standard SMA, LMA and LOW allowed region.
}
\end{figure}

\begin{figure}[p]
\begin{center}
\includegraphics[bb=45 110 540 757,height=0.8\textheight]{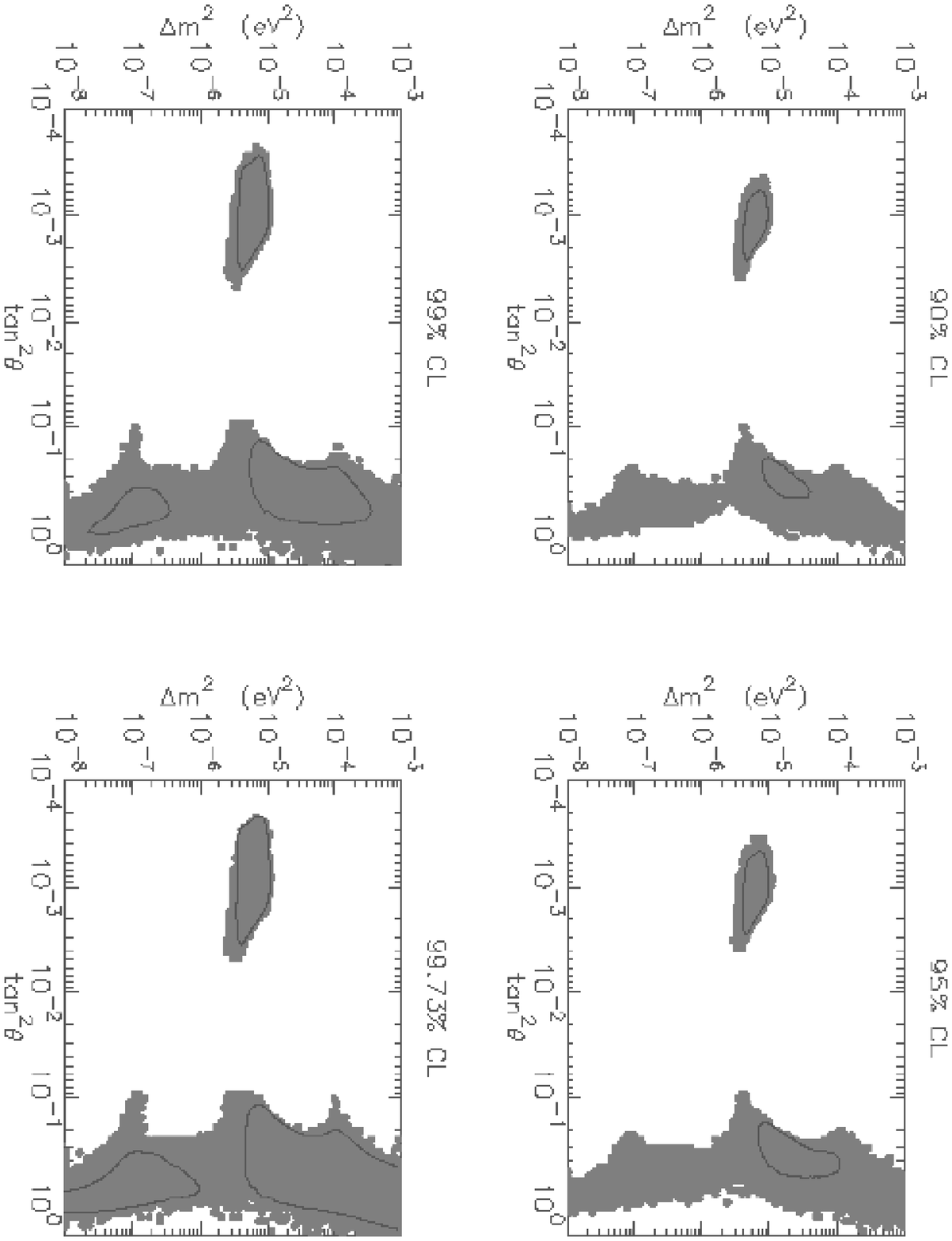}
\end{center}
\caption{ \label{cfi-all-msw}
Allowed 90\%, 95\%, 99\%, 99.73\% confidence level regions
in the
$\tan^2\vartheta$--$\Delta{m}^2$
plane.
The gray areas are the allowed regions with exact frequentist coverage in the
MSW region
obtained without any restriction on the possible
values of $\tan^2\vartheta$ and $\Delta{m}^2$.
The areas enclosed by the solid lines
are the standard SMA, LMA and LOW allowed region.
}
\end{figure}

\begin{figure}[p]
\begin{center}
\includegraphics[bb=45 334 540 757,width=\textwidth]{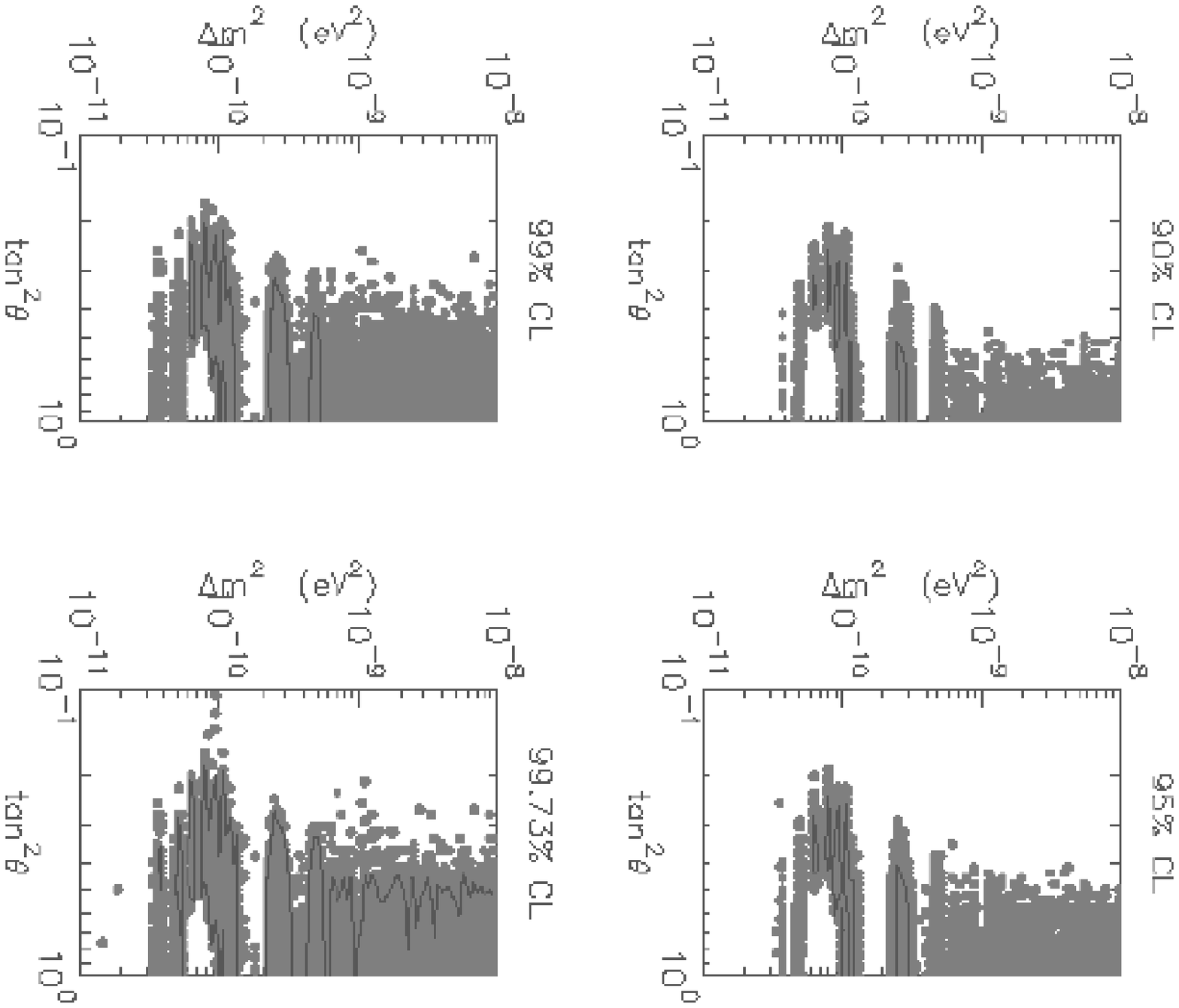}
\end{center}
\caption{ \label{cfi-all-vo}
Allowed 90\%, 95\%, 99\%, 99.73\% confidence level regions
in the
$\tan^2\vartheta$--$\Delta{m}^2$
plane.
The gray areas are the allowed regions with exact frequentist coverage in the
VO region
obtained without any restriction on the possible
values of $\tan^2\vartheta$ and $\Delta{m}^2$.
The areas enclosed by the solid lines
are the standard VO allowed regions.
}
\end{figure}

The results of our calculations
are presented in Figs.~\ref{cfi-sma}--\ref{cfi-all-vo},
where we have depicted the
90\%, 95\%, 99\% and 99.73\% CL regions
(gray areas)
confronted with those obtained
with the standard method based on Eq.~(\ref{DX2})
(areas enclosed by solid lines).

In Fig.~\ref{cfi-sma}
we have restricted the possible values of the neutrino oscillation parameters
in a region around the SMA solution,
where $X^2_{\mathrm{min}}$ lies.
The acceptance interval
for each point on the grid in the parameter space
has been calculated generating
about $6 \times 10^5$ synthetic data sets
(different for different points on the grid, in order to avoid correlations).
One can see that the standard allowed SMA region
is an acceptable approximation of the exact\footnote{
Here the adjective ``exact'' refers
to the method, that produces
confidence intervals with exact coverage.
Obviously our confidence intervals
are approximations of the exact ones,
that would be obtained with an infinitely dense grid
in parameter space and an infinite
set of synthetic random data sets.
}
confidence interval.
This is due to the fact that locally the linear approximation
is rather good,
as we already found in the previous two sections.

In Fig.~\ref{cfi-msw}
we have extended the possible values of the neutrino oscillation parameters
to all the MSW region (\ref{MSW}).
For this figure the number of synthetic data sets
for each point on the grid is about $7 \times 10^4$
(less than in Fig.~\ref{cfi-sma}
because of the larger size of the grid,
that slows down the calculation).
The standard SMA region is still an acceptable approximation
of the exact SMA region,
but the exact LMA and LOW regions are dramatically
larger than the standard ones,
so large that they merge together,
producing a huge allowed region around maximal mixing
($\tan^2\vartheta = 1$).
This is true even at 90\% CL.

Figures \ref{cfi-all-msw} and \ref{cfi-all-vo}
show, respectively, the allowed MSW and VO regions
when there is no restriction on
the possible values of the neutrino oscillation parameters
(the number of synthetic data sets
for each point on the grid is now about $6.5 \times 10^4$).
Again,
one can see that
the standard SMA region is an acceptable approximation
of the exact SMA region,
but the exact LMA, LOW and VO regions are much
larger than the standard ones.

From the results of our calculations we conclude that
the standard method to calculate allowed regions
produces reliable results only locally,
\textit{i.e.}
in the calculation of the allowed region surrounding the global
minimum of $X^2$.
The other allowed regions are dramatically
underestimated by the standard method.

\section{Conclusions}
\label{Conclusions}

We have presented the results of a numerical Monte Carlo
calculation of
the goodness of fit
and the confidence level of the standard allowed regions
for the neutrino oscillation parameters
$\Delta{m}^2$, $\tan^2\vartheta$
obtained from the fit of solar neutrino data.
We have shown that
the standard values of the goodness of fit
and of the confidence level of the allowed regions
are significantly overestimated with the standard method.
This is due to the non-linear dependence of the
neutrino oscillation probability
from the parameters.
The linear approximation,
leading to the standard values
of the goodness of fit
and of the confidence level of the allowed regions,
is valid only locally,
for values of the parameters around a specific MSW solution
(SMA, LMA, LOW).
In the case of the VO solution
the linear approximation is not valid even locally,
because of the strong non-linearity
of the oscillation probability
that causes the existence of several allowed regions close together.

We have also calculated
exact allowed regions
with correct frequentist coverage
using Neyman's method.
The results of these calculations show that
the standard allowed region
around the global minimum of the least-squares function
is a reasonable approximation of the exact one.
On the other hand,
the size of the other regions is dramatically
underestimated in the standard method.
Indeed,
in our calculation
the exact SMA region,
that contains the minimum of the least-squares function,
practically coincides with the standard one.
On the other hand,
the exact LMA and LOW regions
are much larger than the standard ones,
so much that they merge in a huge allowed region
around maximal mixing.
Also the exact allowed VO regions are much larger than the standard ones.

The indications on neutrino mixing coming from solar neutrino
data
are becoming increasingly important for theory and experiment.
Furthermore,
solar neutrino data will soon be enriched by results of new powerful
experiments
(SNO \cite{SNO-99},
Borexino \cite{Borexino-98},
GNO \cite{GNO-99}
and others \cite{vonFeilitzsch-nu2000}).
As we have shown,
the standard statistical analysis of solar neutrino data
can lead to incorrect conclusions
concerning the goodness of fit,
the confidence level of the allowed regions
and
the size of the allowed regions
far from the global minimum of the least-squares function.
Hence,
we believe that it is time to examine critically
the method of statistical analysis of solar neutrino data
and bring it to the level of quality
already attained in other branches of research
in high-energy physics.


\end{document}